\documentclass[a4paper,pre,twocolumn,amsmath,amssymb,longbibliography]{revtex4-1}
\usepackage{bm}
\usepackage{graphicx,color,soul}

\newcommand{\dd}{\mathrm{d}}

\newcommand{\pd}[2]{\frac{\partial #1}{\partial #2}}
\newcommand{\fd}[2]{\frac{\delta #1}{\delta #2}}

\newcommand{\Int}[1]{\int\dd #1\;}
\newcommand{\IInt}[3]{\int_{#2}^{#3}\dd #1\;}
\renewcommand{\vec}[1]{\mathbf #1}
\DeclareMathOperator{\Ei}{Ei}

\newcommand{\al}{\alpha}
\newcommand{\gam}{\gamma}
\newcommand{\eps}{\varepsilon}
\newcommand{\kap}{\kappa}
\newcommand{\lam}{\lambda}

\newcommand{\sig}{\sigma}
\newcommand{\msig}{\bm\sigma}

\newcommand{\kT}{k_\text{B}T}
\newcommand{\tx}{\tau_\text{r}}
\newcommand{\x}{\vec r}

\begin{document}

\title{Coexistence of active Brownian discs: Van der Waals theory and analytical results}

\author{Thomas Speck}
\affiliation{Institut f\"ur Physik, Johannes Gutenberg-Universit\"at Mainz, Staudingerweg 7-9, 55128 Mainz, Germany}


\begin{abstract}
  At thermal equilibrium, intensive quantities like temperature and pressure have to be uniform throughout the system, restricting inhomogeneous systems composed of different phases. The paradigmatic example is the coexistence of vapor and liquid, a state that can also be observed for active Brownian particles steadily driven away from equilibrium. Recently, a strategy has been proposed that allows to predict phase equilibria of active particles [Phys. Rev. E \textbf{97}, 020602(R)(2018)]. Here we elaborate on this strategy and formulate it in the framework of a van der Waals theory for active discs. For a given equation of state, we derive the effective free energy analytically and show that it yields coexisting densities in very good agreement with numerical results. We discuss the interfacial tension and the relation to Cahn-Hilliard models.
\end{abstract}

\maketitle


\section{Introduction}

While pattern formation is typically associated with non-equilibrium, even systems at thermal equilibrium are not necessarily homogeneous. For example, a monocomponent fluid within a container with fixed volume might reduce its global free energy by breaking up into liquid domains surrounded by a gas, slowly coalescing into a single liquid region \emph{coexisting} with the gas occupying the remaining volume. Both liquid and gas have different densities and are separated by an interfacial region of finite width. This phase separation arises when the gain of contact energy outweighs the loss of entropy in the denser phase (the liquid), a simple picture of which is provided by the van der Waals theory~\cite{vanderwaals,rowlinson79,chan83}.

Phase separation \emph{without} cohesive forces is possible in suspensions of repulsive self-propelled particles, which was first reported for run-and-tumble dynamics~\cite{tail08} and later for active Brownian particles~\cite{yaou12,redn13,butt13,sten14,wyso14,digregorio18}. The interplay of excluded volume and directed motion induces a trapping of particles and, therefore, a reduced effective motility that in turn further increases the local density. This mechanism leads to a dynamic instability that has been coined motility-induced phase separation (MIPS)~\cite{cate15}.

The resemblance of MIPS with conventional vapor-liquid coexistence has prompted the search for underlying principles that allow to predict the phase diagram. The ``dynamic route'' to the emerging phase behavior is through studying the hydrodynamic evolution equations of density and higher moments of the particle orientation. These higher moments can be eliminated, leading to an evolution equation for the density alone. To lowest order, one obtains a Cahn-Hilliard-like equation~\cite{spec14,spec15} involving an effective free energy. Pushing this expansion to higher orders, however, introduces further terms that cannot be written as derivatives of a free energy functional anymore~\cite{rapp19,bickmann20,bickmann20a}. Some of these terms had been introduced previously and their effects have been discussed within an ``Active Model B(+)'' for scalar active matter~\cite{sten13,witt14,nard17,tjhung18,caballero18}.

The ``static route'' to the phase behavior is through identifying thermodynamic concepts that can be transferred to active particles~\cite{taka15,paliwal18,wittmann19,hermann19a}. The main ingredient is the (mechanical) pressure exerted by active particles, which has been studied intensively~\cite{mall14,taka14,solo15a,solo15,wink15,spec16a}. However, it has been noted that Maxwell's equal-area construction fails to predict the coexisting densities~\cite{solo15}. Underlying this construction is a free energy, and the equal area constraint translates to an equality of chemical potential. For driven systems such a free energy is absent (although some thermodynamic relations can be established through the work~\cite{spec16}). Solon \emph{et al.} have shown recently that a second condition replacing the equal-area rule can be constructed through transforming non-potential interfacial terms into a total derivative~\cite{solo17,solo18a}. While they have already demonstrated the power of their approach numerically for active discs, here we discuss this strategy in the framework of a mean-field van der Waals theory and show that approximate but accurate expressions for active discs can be obtained analytically in very good agreement with numerical results.


\section{Theory}

\subsection{Conventional coexistence}

In two dimensions, a system at thermal equilibrium is governed by the free energy $F(T,A,N)$, where $A$ is the area (the volume in three dimensions), $N$ is the number of particles, and $T$ is the temperature. The free energy is a homogeneous function, $F=Af(T,\rho)$, with density $\rho=N/A$ and free energy density $f(\rho)$ suppressing the dependence on temperature. Both $A$ and $N$ are extensive variables, and their intensive conjugates are the pressure
\begin{equation}
  p = -\left.\pd{F}{A}\right|_{T,N} = -\pd{(af)}{a} = -f + \rho\pd{f}{\rho}
\end{equation}
and chemical potential
\begin{equation}
  \label{eq:mu:eq}
  \mu = \left.\pd{F}{N}\right|_{T,A} = \pd{f}{\rho} = -a^2\pd{f}{a} = a(f+p),
\end{equation}
where we have employed the reduced area $a=A/N=1/\rho$.

\begin{figure}
  \centering
  \includegraphics{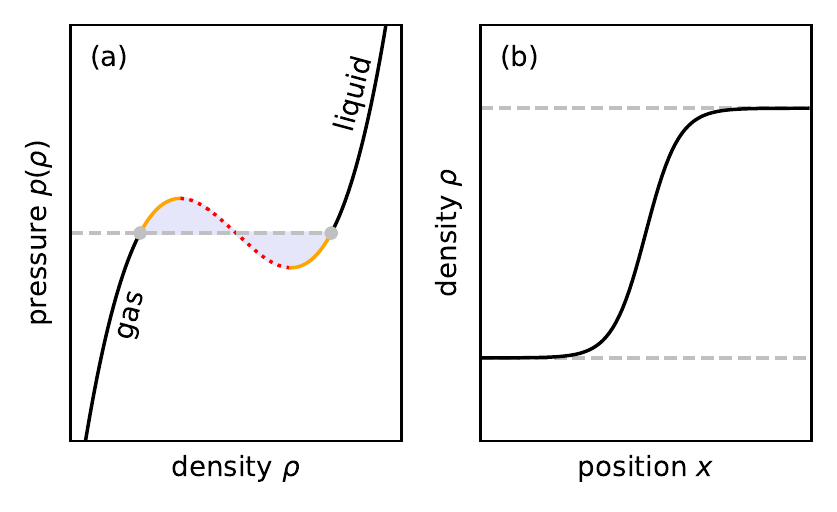}
  \caption{(a)~Sketch of a non-monotonous pressure $p(\rho)$. Coexisting densities (symbols) have the same pressure $p_\ast$, which in equilibrium is found by a Maxwell construction (equal area rule). Continuations of gas and liquid branches (orange) are metastable, while densities along the dotted red line are unstable. (b)~Density profile $\rho(x)$ with coexisting densities (dashed lines).}
  \label{fig:maxwell}
\end{figure}

In an inhomogeneous system at coexistence, the density $\rho(\x)$ is non-uniform while intensive pressure and chemical potential are constant throughout. As equations of state, $p=p(\rho)$ and $\mu=\mu(\rho)$ become non-monotonous below a critical temperature, cf. Fig.~\ref{fig:maxwell}(a), the most famous example for which is arguably van der Waals' phenomenological equation of state~\cite{vanderwaals,rowlinson79}. At intermediate densities attractive forces reduce the pressure before it rises again due to the strong repulsion as particles come in close range. Although the coexisting densities are solely determined by the bulk free energy density $f$, within the interfacial region forces due to the density gradient need to be compensated. To lowest order, this can be captured by a contribution
\begin{equation}
  \label{eq:F:nabla}
  \mathcal F_\nabla = \Int{^2\x} \frac{1}{2}\kT\kap(\rho)|\nabla\rho|^2
\end{equation}
to the free energy penalizing gradients with coefficient $\kap$, and $\mathcal F=\mathcal F_\nabla+\Int{^2\x}f(\rho(\x))$ is called the Ginzburg-Landau free energy~\cite{hohenberg15}.

\subsection{Active Brownian particles}

We consider interacting active Brownian particles moving in two dimensions. Particles are propelled with constant speed $v_0$ along their orientation. This directed motion breaks detailed balance, keeping the system in a non-equilibrium steady state. The particle current can be expressed as
\begin{equation}
  \label{eq:j}
  \vec j = v_0\vec p + \mu_0\nabla\cdot\msig_\text{IK} - D_0\nabla\rho,
\end{equation}
where $\msig_\text{IK}$ is the symmetric Irving-Kirkwood stress tensor due to interparticle forces~\cite{irvi50} and the diffusion coefficient $D_0=\kT\mu_0$ is related to the bare mobility $\mu_0$ through the temperature $T$ with Boltzmann's constant $k_\text{B}$. Here both the density $\rho$ and the polarization $\vec p$ enter, see Ref.~\citenum{speck20} for a derivation. Even though particle orientations diffuse independently, the polarization can be non-zero. In the following, we take the perspective that the polarization can be related to the divergence of an active stress, $v_0\vec p=\mu_0\nabla\cdot\msig_\text{A}$, where
\begin{equation}
  \label{eq:sig:act}
  \msig_\text{A} = -v_0\tx\left[\frac{v}{2\mu_0}\rho\bm 1 + \frac{v}{\mu_0}\vec Q - \kT\nabla\vec p\right]
\end{equation}
with the symmetric and traceless nematic tensor $\vec Q(\x)$. Here we have applied the ``force closure'' to eliminate the correlations between orientations and forces~\cite{speck20}, which introduces an effective speed $v=v(\rho)$ depending on the local density. The exact form of $v(\rho)$ depends on details of the interactions. We can thus write the particle current $\vec j=\mu_0\nabla\cdot\msig$ with the total stress tensor $\msig=-\kT\rho\bm 1+\msig_\text{IK}+\msig_\text{A}$.

\subsection{Mechanical equilibrium and effective chemical potential}

At coexistence the particle current vanishes, which yields the condition of mechanical equilibrium $\nabla\cdot\msig=0$. We now restrict our attention to a planar average interface separating a dense and a dilute region. We orient the coordinate system so that the interface normal points along $\vec e_x$ with polarization $\vec p=p_x(x)\vec e_x$ and density $\rho=\rho(x)$. The normal component of the total stress reads
\begin{equation}
  \label{eq:sig}
  \sig_{xx} = -p_\perp - \frac{v_0\tx}{\mu_0}(vQ_{xx} - D_0\partial_xp_x).
\end{equation}
In each of the bulk phases, $Q_{xx}=0$ and $p_x=0$ and the second term vanishes. The first term $p_\perp\to p$ then becomes the isotropic bulk pressure
\begin{equation}
  \label{eq:p}
  p(\rho) = \kT\rho + p_\text{IK}(\rho) + \frac{v_0\tx v(\rho)}{2\mu_0}\rho,
\end{equation}
which only depends on the density~\cite{solo15}.

A vanishing divergence $\partial_x\sig_{xx}=0$ requires that $\sig_{xx}=-p_\ast$ is constant throughout and thus $p(\rho_+)=p(\rho_-)=p_\ast$, where $\rho_\pm$ are the coexisting bulk densities far away from the interface [Fig.~\ref{fig:maxwell}(b)]. A single condition is not sufficient since we do not know the value of $p_\ast$. In passive systems, a necessary second condition is obtained through equating the chemical potentials in both phases. In Ref.~\citenum{solo17}, Solon \emph{et al.} describe how an integrating factor can be used to transform the stress profile [Eq.~\eqref{eq:sig}] into a total derivative, which allows to define a function that takes on the role of the chemical potential.

Here is how it works. Let us consider the integral
\begin{equation}
  \IInt{x}{x_-}{x_+} \sig_{xx}\partial_xa = -p_\ast(a_+ - a_-)
\end{equation}
through the interface, where $x_\pm$ are positions within the two bulk phases. We have used that the stress has to be constant, and $a(x)$ is an unknown function with $a_\pm=a(x_\pm)$. For a second relation, we now consider the expression Eq.~\eqref{eq:sig} and write $\sig_{xx}\partial_xa=\partial_x(a\psi)$ with yet another function $\psi(x)$ to be determined. Now the integral becomes
\begin{equation}
  \IInt{x}{x_-}{x_+} \sig_{xx}\partial_xa = a_+\psi_+ - a_-\psi_-
\end{equation}
with $\psi_\pm=\psi(x_\pm)$. Hence, equating both integrals leads to the condition $\mu(x_+)=\mu(x_-)$ with function $\mu\equiv a(\psi+p)$. Clearly, $\mu$ looks like a chemical potential with free energy density $\psi$, cf. Eq.~\eqref{eq:mu:eq}. This interpretation endows the function $a(x)$ with the meaning of an effective local area. However, it is important to note that $a$ is no longer given as $1/\rho$.

\subsection{Approximating the stress}

To determine the integrating factor and establish its relation to the actual density, we make several approximations to Eq.~\eqref{eq:sig}. First, we assume $p_\perp\approx p$. For the second term, we need expressions for $p_x$ and $Q_{xx}$. For the polarization, we employ the force closure $\nabla\cdot\msig_\text{IK}\propto-\vec p$ capturing the force imbalance due to the effective alignment of particles within the interfacial region~\cite{spec15}. Hence, $j_x\approx vp_x-D_0\partial_x\rho=0$ with the same effective speed $v(\rho)$ as in Eq.~\eqref{eq:sig:act}. Finally, this relation allows to obtain a closed expression for the nematic tensor with normal component
\begin{equation}
  Q_{xx} \approx -\frac{1}{16}\tx\partial_x(vp_x) \approx -\frac{1}{16}D_0\tx\partial_x^2\rho.
  \label{eq:Q}
\end{equation}
Plugging this result together with $p_x=D_0\partial_x\rho/v(\rho)$ into Eq.~\eqref{eq:sig}, we obtain
\begin{multline}
  \label{eq:sig:a}
  \frac{v_0\tx}{D_0}(vQ_{xx} - D_0\partial_xp_x) \\ = -v_0\tx\left(\frac{v\tx}{16}+\frac{D_0}{v}\right)\partial_x^2\rho + v_0\tx\frac{D_0v'}{v^2}|\partial_x\rho|^2 \\ \equiv -\kap(\rho)\partial_x^2\rho + \lam(\rho)|\partial_x\rho|^2,
\end{multline}
which defines two functions $\kap(\rho)$ and $\lam(\rho)$ that depend only on the local density. The prime denotes the derivative with respect to the density, $v'=dv/d\rho$. The right-hand side of Eq.~\eqref{eq:sig:a} can only be expressed as the functional derivative of an interfacial free energy if $2\lam+\kap'=0$. This relation follows from
\begin{equation}
  \fd{\mathcal F_\nabla}{\rho} = \frac{1}{2}\kT\kap'(\rho)|\nabla\rho|^2 - \kT\nabla\cdot[\kap(\rho)\nabla\rho]
\end{equation}
after one integration by parts and using $\nabla\cdot[\kap\nabla\rho]=\kap'|\nabla\rho|^2+\kap\nabla^2\rho$. In general, $2\lam+\kap'\neq0$ and the interfacial stress cannot be expressed through a free energy.

\subsection{Integrating factor and effective free energy}

To proceed we follow Ref.~\citenum{solo17}. Consider the derivative
\begin{equation}
  \partial_x\left(\frac{\kap a'}{2}|\partial_x\rho|^2\right) = \left[\frac{\kap'a'+\kap a''}{2}|\partial_x\rho|^2+\kap a'\partial_x^2\rho\right]\partial_x\rho,
  \label{eq:map}
\end{equation}
which reduces to Eq.~\eqref{eq:sig:a} (times $\partial_xa$) if we set
\begin{equation}
  \kap a'' = -(2\lam+\kap')a'.
\end{equation}
This constitutes a differential equation that determines $a=a(\rho)$ as a function of the local density $\rho$. Plugging in the expressions for $\kap$ and $\lam$ from Eq.~\eqref{eq:sig:a}, we find $2\lam+\kap'=\kap v'/v$ and thus $a''/a'=-v'/v$. This can be integrated once to yield $a'=c/v$ with integration constant $c$.

Since the bulk pressure involves no gradients, we can write $p=-\pd{(a\phi)}{a}$ again in analogy with the pressure in passive suspensions. Putting everything together, we obtain
\begin{equation}
  \sig_{xx}\partial_xa = \partial_x\left(a\phi + \frac{\kT\kap a'}{2}|\partial_x\rho|^2\right) = \partial_x(a\psi),
\end{equation}
from which we read off
\begin{equation}
  \label{eq:psi}
  \psi = \frac{\kT\kap_\text{eff}}{2}|\partial_x\rho|^2 + \phi
\end{equation}
with $\kap_\text{eff}(\rho)\equiv\kap a'/a$. The function $\psi=\psi(\rho,\partial_x\rho)$ takes on the form of a Ginzburg-Landau free energy density with bulk free energy density $\phi(\rho)$ plus a term that penalizes sharp density gradients. Away from the interface in the bulk phases the gradients vanish and equality of the ``chemical potential'' $\mu$ reduces to the condition
\begin{equation}
  \label{eq:mu}
  a_+(\phi_+ + p_\ast) = a_-(\phi_- + p_\ast).
\end{equation}
To proceed further we need an expression for the bulk pressure $p=p(\rho)$ from which we can determine $\phi(\rho)$ and then exploit both conditions on the pressure and Eq.~\eqref{eq:mu} to find the coexisting densities.


\section{Active hard discs}

\subsection{Equation of state}

We now focus on the most studied variant of active Brownian particles interacting through short-range repulsive pair forces. In this case the density-dependent speed follows a linear decrease~\cite{sten13,fischer19}
\begin{equation}
  \label{eq:v}
  v(\rho) = v_0(1-\rho/\rho_0)
\end{equation}
surprisingly well up to high densities close to the ``jamming'' density $\rho_0$, beyond which $v=0$. Solving $a'=c/v$, we find
\begin{equation}
  a(\rho) = -a_0\ln(1-\rho/\rho_0), \quad \rho/\rho_0 = 1-e^{-a/a_0}
\end{equation}
with arbitrary constant $a_0$.

\begin{figure}[t]
  \centering
  \includegraphics{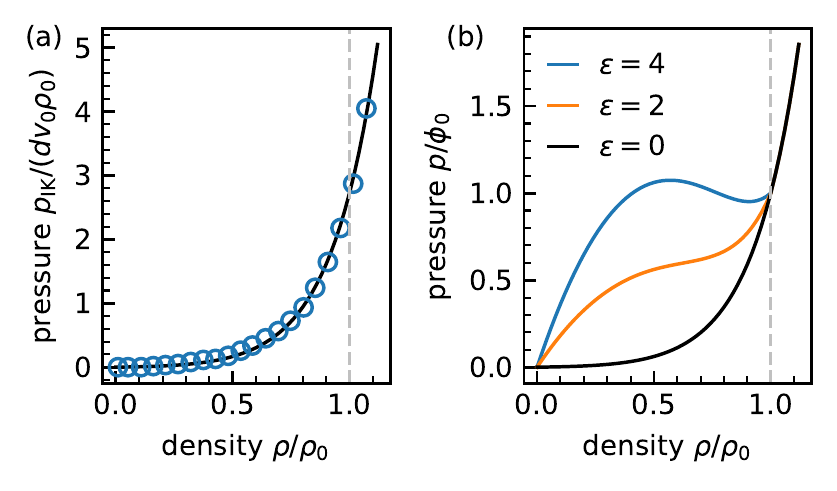}
  \caption{(a)~The pressure $p_\text{IK}$ due to interparticle forces as a function of density $\rho$ taken from Ref.~\citenum{solo15}. The line is a fit to Eq.~\eqref{eq:p:IK}. (b)~The total bulk pressure $p(\rho)$ [Eq.~\eqref{eq:eos}] as a function of density for three values of $\eps$. For densities $\rho>\rho_0$ the effective speed $v=0$ is zero and all curves collapse onto $p_\text{IK}$.}
  \label{fig:eos}
\end{figure}

Upon inserting Eq.~\eqref{eq:v}, the first contribution to the pressure [Eq.~\eqref{eq:p}] becomes
\begin{equation}
  p_\text{A} = \kT\rho + \frac{v_0\tx v(\rho)}{2\mu_0}\rho \approx \frac{v_0^2\tx\rho_0}{2\mu_0}z(1-z)
\end{equation}
with scaled density $z\equiv\rho/\rho_0$. In the following we neglect the first term $\kT\rho$ due to diffusive momentum transfer, which is much smaller than the active contribution. For the remaining Irving-Kirkwood contribution, Ref.~\citenum{solo15} argues that an exponential increase works well, and we choose
\begin{equation}
  \label{eq:p:IK}
  p_\text{IK} = \frac{v_0\rho_0}{\mu_0}\al ze^{\xi(z-1)}
\end{equation}
with two fit parameters, dimensionless $\xi$ and length $\al$, which are assumed to be independent of the speed. This function is shown in Fig.~\ref{fig:eos}(a) fitted to the numerical data provided in Ref.~\citenum{solo15} with $\xi\simeq4.20$ and $\al\simeq2.72d$ with particle diameter $d$. We thus have constructed an approximate closed form
\begin{equation}
  \frac{p(z)}{\phi_0} = 
  \begin{cases}
    \eps z(1-z)+ze^{\xi(z-1)} & z \leqslant 1 \\
    ze^{\xi(z-1)} & z > 1
  \end{cases}
  \label{eq:eos}
\end{equation}
for the equation of state $p=p(z)$. Here we have defined the prefactor $\phi_0\equiv\kT\rho_0v_0\al/D_0$ and the scaled persistence length
\begin{equation}
  \eps \equiv \frac{v_0\tx}{2\al}
\end{equation}
as the dimensionless control parameter.

In Fig.~\ref{fig:eos}(b), the pressure $p/\phi_0$ is plotted for several values of $\eps$. We see that beyond a critical $\eps_\text{c}$ the pressure becomes non-monotonous as discussed in Fig.~\ref{fig:maxwell}. The reason is that the active pressure $p_\text{A}$ drops as particles become trapped and the effective speed $v$ is reduced, the same effect that is achieved through attractive forces in the van der Waals theory. The relation $p=p(\rho)$ is sufficient to determine the mean-field spinodal through $p'=0$, \emph{i.e.}, the line within which a homogeneous system is linearly unstable. The spinodal terminates in the critical point, which is found combining the conditions $p'=0$ and $p''=0$ yielding
\begin{gather}
  \eps_\text{c} = \frac{\xi}{2}(2+\xi z_\text{c})e^{\xi(z_\text{c}-1)}, \\
  z_\text{c} = \frac{\xi-2+\sqrt{\xi^2+12\xi+20}}{4\xi}
\end{gather}
as a function of $\xi$. Specifically, for $\xi=4.2$ we find $z_\text{c}\simeq0.69$ and $\eps_\text{c}\simeq2.79$. In terms of the P\'eclet number $\text{Pe}=3v_0\tx/d=6\eps\al/d$ the critical value is $\text{Pe}_\text{c}\simeq45.5$, which is close to the estimate $\text{Pe}_\text{c}\simeq40$ found in Ref.~\citenum{sieb18} for a slightly different system.

\subsection{Bulk free energy}

\begin{figure}[b!]
  \centering
  \includegraphics{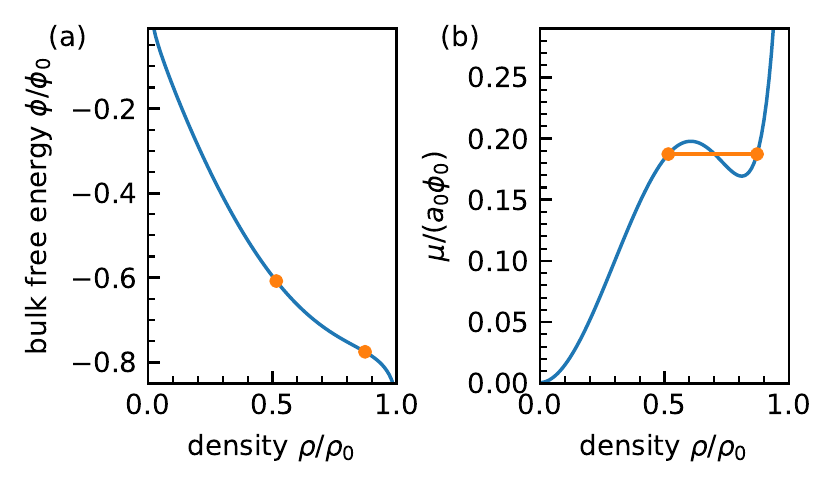}
  \caption{(a)~Effective bulk free energy $\phi(\rho)$ from Eq.~\eqref{eq:phi} for $\eps=3.2$ and $\xi=4.20$. Indicated are the coexisting densities (symbols). Since the chemical potential is not given by the derivative, the tangents do not have the same slope. (b)~The effective chemical potential $\mu(\rho)$ is equal for both coexisting densities (symbols).}
  \label{fig:phi}
\end{figure}

The bulk free energy can be determined through integrating the equation of state
\begin{equation}
  \phi(a) = -\frac{1}{a}\IInt{s}{0}{a}p(\rho(s))
\end{equation}
with irrelevant integration constant. Note that
\begin{equation}
  \phi'(\rho) = a'\pd{\phi}{a} = -\frac{a'}{a}(\phi+p),
\end{equation}
which in general is different from $\mu=a(\phi+p)$ except for $a=1/\rho$. Let us make the substitution $z=\rho/\rho_0=1-e^{-a/a_0}$ leading to the integral
\begin{equation}
  \IInt{s}{0}{a}p(s) = a_0\IInt{z}{0}{\rho/\rho_0} \frac{p(z)}{1-z}.
\end{equation}
This integral can be performed with result
\begin{equation}
  \frac{\phi(z)}{\phi_0} = \frac{1}{\ln(1-z)} \left(\frac{1}{2}\eps z^2-\Ei(\xi z-\xi)-\frac{1}{\xi}e^{\xi z-\xi}\right),
  \label{eq:phi}
\end{equation}
where $\Ei(x)$ is the exponential integral. The function $\phi/\phi_0$ is plotted in Fig.~\ref{fig:phi}(a) for one value of $\eps>\eps_\text{c}$ together with the coexisting densities. Clearly, the coexisting densities are not related by a common tangent, which is a consequence of the fact that $a\neq1/\rho$ is not the reduced area. However, the effective chemical potential $\mu(\rho)$ is equal in both phases by construction [Fig.~\ref{fig:phi}(b)].

\subsection{Phase diagram}

\begin{figure}[b!]
  \centering
  \includegraphics{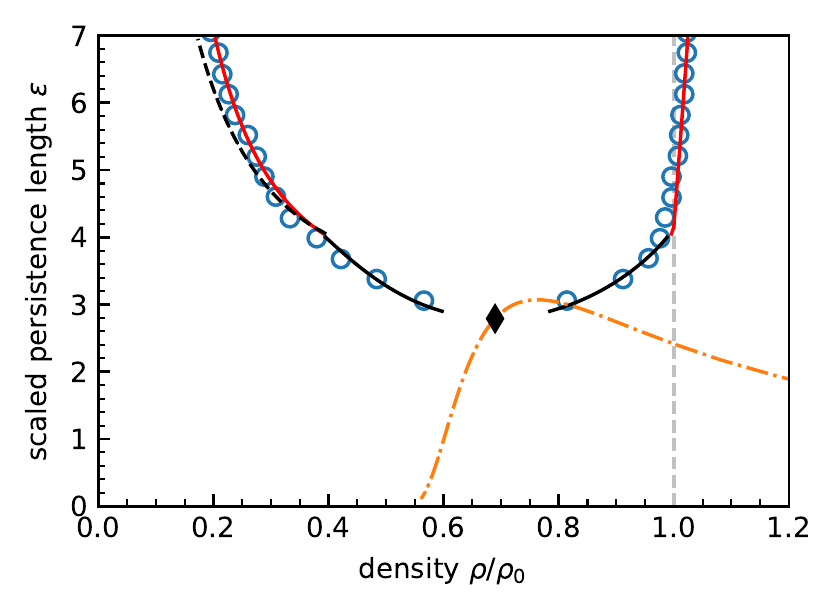}
  \caption{Coexisting densities. Circles show numerical results from Ref.~\citenum{solo15}. The solid black line is the prediction from the analytical expressions for pressure and effective chemical potential, which is only available for $\rho\leqslant\rho_0$. The black dashed line shows the isobar $p=\phi_0$ and solid red lines are coexisting densities using $p/\phi_0=1+b(\mu-\mu_1)$ with $b=0.1$ for $\mu>\mu_1$. The dash-dotted line is the critical point as a function of $\xi$ (moving to smaller $\rho/\rho_0$ as $\xi$ increases) with the diamond indicating the critical point for the corresponding $\xi\simeq4.20$.}
  \label{fig:binodal}
\end{figure}

Now that we have $\mu=a(\phi+p)$ we obtain the coexisting densities $\rho_\pm$ numerically through root finding. The resulting values are plotted in Fig.~\ref{fig:binodal} as a function of $\eps$ (solid line). Starting from the critical point, the binodal opens until the larger density $\rho_+$ reaches $\rho_0$ at $\eps_\times\simeq 4$. At this point the effective speed $v(\rho_+)=0$ in the dense liquid vanishes and there is a crossover caused by the fact that the active liquid branch ends at the chemical potential
\begin{equation}
  \mu_1(\eps) \equiv \lim_{z\to1^-} \mu(z) = a_0\phi_0\left(\gam_\text{e}+\frac{1}{\xi}+\ln\xi-\frac{\eps}{2}\right),
\end{equation}
which remains finite although $a$ diverges as $z\to 1$. Here, $\gam_\text{e}\simeq0.577$ is the Euler-Mascheroni constant. That the active liquid branch ends at a finite $\mu_1$ is an artifact of approximating the speed through Eq.~\eqref{eq:v} with $v=0$ for $z>1$. Improving the functional form of $v(z)$ so that it decays fast for $z>1$ while remaining non-zero will remove this artifact but is more difficult to handle analytically.

\begin{figure}[t]
  \centering
  \includegraphics{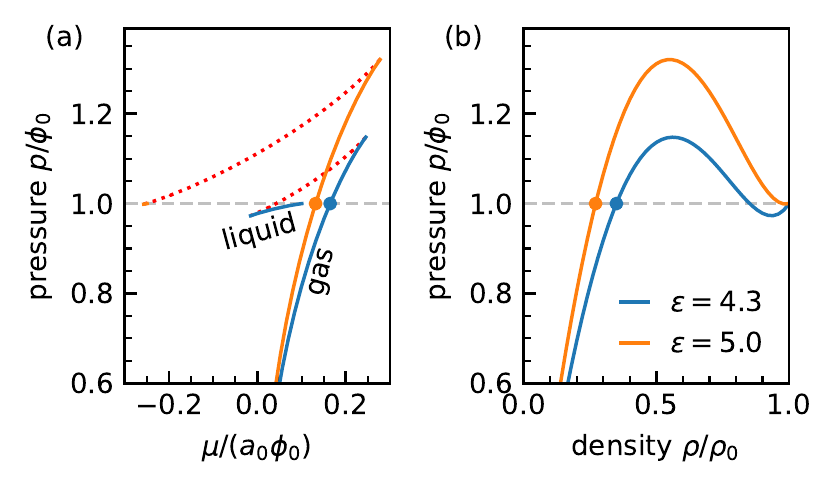}
  \caption{Coexistence at large $\eps>\eps_\times$. (a)~Pressure as function of chemical potential $\mu$ for $\eps=4.3$ and $\eps=5$. The unstable densities are indicated as dotted red lines. The active liquid branch ends at $\mu_1$ for $\eps=4.3$ and is absent for $\eps=5$. Symbols indicate coexistence values for the chemical potential using $p(\mu)=\phi_0$ for $\mu>\mu_1$. (b)~Pressure as function of density for the same values of $\eps$. Symbols indicate the gas densities $\rho_-$.}
  \label{fig:intensive}
\end{figure}

Keeping the simplified form of $v(z)$ and increasing $\eps$ beyond $\eps_\times$, the active liquid branch no longer crosses the active gas branch as shown in Fig.~\ref{fig:intensive}(a). To proceed, we require a continuation of the pressure $p(\mu)$ for $\mu>\mu_1$. The simplest approximation is to assume that the pressure $p(\mu)=\phi_0$ is independent of $\mu$ with $\rho_+=\rho_0$. The gas density $\rho_-$ is then obtained from the isobar $p(\rho_-)=\phi_0$ [Fig.~\ref{fig:intensive}(b)] and shown in Fig.~\ref{fig:binodal} as dashed line. Also plotted are the numerical values for $\rho_\pm$ taken from Ref.~\citenum{solo15}, which agree well with our predictions without any fitting (except $\al$ and $\xi$ entering the pressure $p_\text{IK}$). The agreement can be improved further using $p(\mu)/\phi_0=1+b(\mu-\mu_1)$ for $\mu>\mu_1$ at the expense of another fit parameter $b$.

\section{Discussion}

\subsection{Interfacial tension}

Maintaining the interface between the two phases requires forces and increases the free energy, which is captured by the interfacial tension $\gam$. There are (at least) two routes to determine this quantity, the mechanical route and the thermodynamic route, which lead to the same result in equilibrium.

We first follow the mechanical route. For our planar interface and coordinate system, the off-diagonal components of the stress vanish, $\sig_{xy}=\sig_{yx}=0$. The tangential stress component
\begin{equation}
  \label{eq:sig:tan}
  \sig_{yy} = -p_\parallel + \frac{v_0\tx}{\mu_0}vQ_{xx},
\end{equation}
however, is non-zero and not restricted by mechanical equilibrium. There can thus be tangential forces within the interfacial region (see Ref.~\citenum{zakine20} for a discussion of these forces in contact with solids). Here we have used that the nematic tensor is traceless, $Q_{yy}=-Q_{xx}$. According the Kirkwood and Buff~\cite{kirk49}, the interfacial tension due to a non-uniform mechanical stress can be expressed as
\begin{equation}
  \gam_\text{KB} = \IInt{x}{x_-}{x_+} [\sig_{yy}(x)-\sig_{xx}].
\end{equation}
We assume that the difference $p_\perp-p_\parallel$ is negligible. Plugging in Eqs.~\eqref{eq:sig} and~\eqref{eq:sig:tan} together with Eq.~\eqref{eq:Q} for the nematic tensor leads to
\begin{equation}
  \gam_\text{KB} = -\frac{\kT v_0\tx^2}{8}\IInt{x}{x_-}{x_+} v\partial_x^2\rho,
\end{equation}
whereby $\Int{x}\partial_xp=0$. Using $\partial_x(v\partial_x\rho)=v'|\partial_x\rho|^2+v\partial_x^2\rho$ we thus obtain
\begin{equation}
  \gam_\text{KB} = \kT\IInt{x}{x_-}{x_+} \kap_\text{KB}(\rho)|\partial_x\rho|^2
\end{equation}
with $\kap_\text{KB}(\rho)\equiv v_0\tx^2v'(\rho)/8$. Specifically, for active hard discs with speed Eq.~\eqref{eq:v}, we find $\kap_\text{KB}=-(v_0\tx)^2/(8\rho_0)<0$. This quantity is negative in agreement with numerical simulations~\cite{bial15,patch18}, a fact that is somewhat counter-intuitive and has motivated alternative definitions of the interfacial tension~\cite{hermann19,omar20}.

A second route to the interfacial tension is through the dependence of the (effective) free energy on the length of the interface $\ell$. Consider an area-preserving transformation $\x\to\vec h\cdot\x$ with
matrix
\begin{equation}
  \vec h = \left(\begin{array}{cc}
    1-h_\parallel & 0 \\ 0 & 1+h_\parallel    
  \end{array}\right).
\end{equation}
While the bulk free energy is not affected, the interfacial free energy $\mathcal F_\nabla$ transforms as
\begin{equation}
  \ell\Int{x}|\partial_x\rho|^2 \to \ell(1+h_\parallel)\Int{x}\frac{1}{1-h_\parallel}|\partial_x\rho|^2.
\end{equation}
To linear order we have $\delta\mathcal F_\nabla=\gam\ell h_\parallel$ reading off the interfacial tension~\cite{evan79}
\begin{equation}
  \gam = \kT\IInt{x}{x_-}{x_+} \kap_\text{eff}(\rho)|\partial_x\rho|^2
\end{equation}
using
\begin{equation}
  \frac{1+h_\parallel}{1-h_\parallel} = (1+h_\parallel)(1+h_\parallel+\cdots) = 1+2h_\parallel+\cdots
\end{equation}
Clearly, $\kap_\text{eff}(\rho)>0$ is positive, a function of density, and does not agree with $\kap_\text{KB}$. In equilibrium, the free energy determines both the reversible work and fluctuations. The disagreement between mechanical and (effective) thermodynamic route to the interfacial tension is \emph{not} an inconsistency and simply shows that such a free energy does not exist for active particles. Other effective thermodynamic treatments also obtain a positive interfacial tension~\cite{hermann19}. Of course, one can associate an effective (isotropic) stress $\sig_\text{eff}$ with the effective free energy density $\psi$, but obviously that leads to the same $\gam$.

\subsection{Cahn-Hilliard models}

\begin{figure}[b!]
  \centering
  \includegraphics{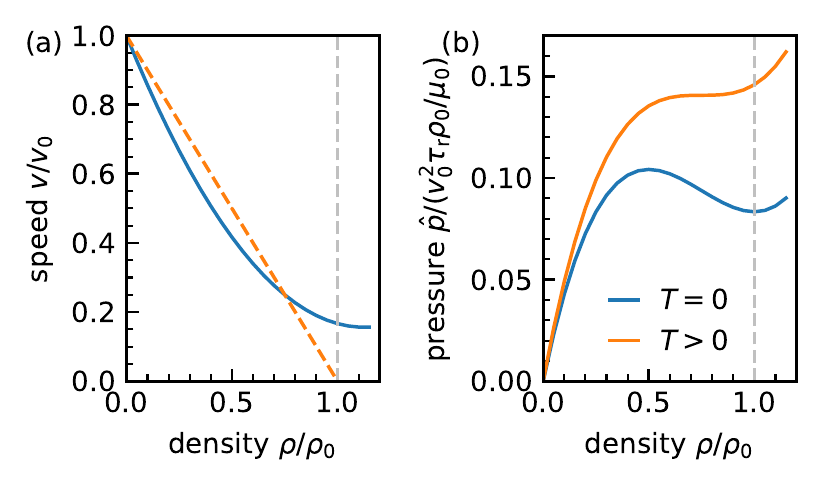}
  \caption{(a)~Effective speed $v$ entering the hydrodynamic equations (dashed line) and $\hat v$ (solid line) determining the pressure $\hat p$. (b)~The pressure $\hat p$ [Eq.~\eqref{eq:p:cahn}] for $T=0$ and the critical temperature at which the minimium vanishes. Curves are only meaningful for $\rho\leqslant\rho_0$.}
  \label{fig:cahn}
\end{figure}

Another route to construct a free energy is through the hydrodynamic evolution equations for density, $\partial_t\rho+\nabla\cdot\vec j=0$, and polarization,
\begin{equation}
  \partial_t\vec p = \frac{\mu_0}{v_0\tx}\nabla\cdot\msig_\text{A} - \frac{1}{\tx}\vec p.
\end{equation}
To achieve closure, we assume $\vec Q\approx 0$ and $\nabla^2\vec p\approx 0$ and apply the force closure $\mu_0\nabla\cdot\msig_\text{IK}=-(v_0\rho/\rho_0)\vec p$ in both evolution equations~\cite{bial13}. Note that this implies that $\msig_\text{IK}$ does not enter explicitly and interactions are only accounted for through $v=v(\rho)$. The polarization can be eliminated systematically using an expansion in the distance to the line marking linear instability (the spinodal), which to lowest order yields an evolution equation for the density that can be written as a Cahn-Hilliard equation with current~\cite{spec14,spec15}
\begin{equation}
  \vec j = -\nabla\fd{\hat{\mathcal F}}{\rho} = \mu_0\nabla\hat\sig.
\end{equation}
Departing from the usual interpretation as a chemical potential, and in the light of our discussion here, we interpret the right hand side as an isotropic stress $\hat\sig\bm 1$. This stress governs the evolution of \emph{small} deviations from a uniform density with constant $\kap$ and $\lam=0$. The pressure can be cast into
\begin{equation}
  \label{eq:p:cahn}
  \hat p = \kT\rho + \frac{v_0^2\tx}{2\mu_0}\left(1-\frac{3}{2}z+\frac{2}{3}z^2\right)\rho.
\end{equation}
In analogy with Eq.~\eqref{eq:p}, the second term looks like an active pressure but with an effective speed $\hat v=v_0(1-\tfrac{3}{2}z+\tfrac{2}{3}z^2)$ different from $v=v_0(1-z)$. In Fig.~\ref{fig:cahn}(a), both speeds are plotted. The pressure $\hat p$ shown in Fig.~\ref{fig:cahn}(b) is non-monotonous for sufficiently small temperatures and thus allows coexistence of two densities. It is clear that it is not the mechanical pressure exerted on walls but a surrogate generated by the hydrodynamic equations. It captures the MIPS mechanism for the dynamic instability but not the balance of active and interparticles forces required for making quantitatively accurate predictions for the coexisting densities.


\section{Conclusions}

We have argued that it is useful to think about active Brownian particles in terms of a stress tensor that is composed of a contribution due to interparticle forces and an active stress tensor $\msig_\text{A}$ [Eq.~\eqref{eq:sig:act}]. Its divergence $\nabla\cdot\msig_\text{A}=(v_0/\mu_0)\vec p$ generates an ``active force'' along the polarization. For a planar interface, balancing this active force with the density gradient yields the approximation
\begin{equation}
  \sig_{xx} \approx -p(\rho) + \kT\kap(\rho)\partial_x^2\rho - \kT\lam(\rho)|\partial_x\rho|^2
  \label{eq:sig:appr}
\end{equation}
with bulk pressure $p(\rho)$ [Eq.~\eqref{eq:p}] and functions $\kap(\rho)$ and $\lam(\rho)$ defined through Eq.~\eqref{eq:sig:a}. The stress thus only depends on the local density and its derivatives. A similar expression has been discussed previously within ``Active Model B'' as an effective chemical potential~\cite{sten13,witt14}, while here we identify it with the (mechanical) stress. This distinction facilitates the identification with the conventional notions and expressions of pressure and chemical potential through a mapping to an effective free energy density [Eq.~\eqref{eq:psi}]. Such a mapping is always possible in one dimension through introducing an integrating factor $\partial_xa$, which turns the non-potential second term in Eq.~\eqref{eq:sig:appr} into a total derivative, cf. Eq.~\eqref{eq:map}. The function $a(\rho)$ takes on the role of a generalized order parameter that is, however, different from the actual reduced area, $a\neq1/\rho$.

In analogy with the van der Waals theory for vapor-liquid equilibrium, from a modeled or measured equation of state $p=p(\rho)$ we can extract a free energy density $\phi(\rho)$ that allows to obtain the chemical potential $\mu$. Phase coexistence is then determined by the same relations as in conventional equilibrium theory but involving $a(\rho)$ instead of $1/\rho$. For active hard discs, we employ an equation of state [Eq.~\eqref{eq:eos}] proposed previously~\cite{solo15,solo18a}. Following the outlined strategy, we derive a closed expression for $\phi$ [Eq.~\eqref{eq:phi}] and predict coexisting densities $\rho_\pm$ that are in very good agreement with the numerically measured coexisting densities without any further adjustment, cf. Fig~\ref{fig:binodal}. However, active Brownian particles are still driven away from thermal equilibrium and $\phi$ is not the free energy of a driven system, but the free energy of another--passive--system that yields the same density profile as the driven system. The breakdown of this analogy can be clearly seen for the interfacial tension, for which mechanical and thermodynamic route yield different results in contrast to equilibrium.


\begin{acknowledgments}
  I acknowledge financial support by the Deutsche Forschungsgemeinschaft (DFG) within the priority program SPP 1726 (Grant No. 254473714).
\end{acknowledgments}


%
  
\end{document}